\documentclass[traditabstract]{aa} 
\usepackage{epsfig}
\usepackage{graphicx}
\usepackage{txfonts}
\begin{document}
\title{Source Counts at 15 microns from the AKARI NEP Survey }

   \subtitle{}

   \author{C.P.~Pearson
          \inst{1,2,7}
          \and
          S. Oyabu\inst{3}
          \and
          T. Wada\inst{3}
          \and
          H. Matsuhara\inst{3}
          \and
          H.M. Lee\inst{4}
          \and
          S.J. Kim\inst{4}
          \and
          T. Takagi\inst{3}
          \and
          T. Goto\inst{5,6}
          \and
          M.S. Im\inst{4}
          \and
          S. Serjeant\inst{7}
          \and
          M.G. Lee\inst{4}
          \and
          J.W. Ko\inst{4}
          \and
          G.J. White\inst{7}
          \and
          O. Ohyama\inst{8}
          }

   \institute{Space Science and Technology Department, CCLRC Rutherford Appleton Laboratory, Chilton, Didcot, Oxfordshire OX11 0QX, UK\\
              \email{chris.pearson@stfc.ac.uk}
         \and
             Department of Physics, University of Lethbridge, 4401 University Drive, Lethbridge, Alberta T1J 1B1, Canada\\
         \and
             Institute of Space and Astronautical Science, Yoshinodai 3-1-1, Sagamihara, Kanagawa 229 8510, Japan\\
         \and
             Department of Physics and Astronomy, Seoul National University, Shillim-Dong, Kwanak-Gu, Seoul 151-742, Korea\\
          \and
             Institute for Astronomy, University of Hawaii, 2680 Woodlawn Drive, Honolulu, HI, 96822, USA\\
         \and
             National Astronomical Observatory, 2-21-1 Osawa, Mitaka, Tokyo, 181-8588,Japan\\
          \and
             Astrophysics Group, Department of Physics, The Open University, Milton Keynes, MK7 6AA, UK\\
          \and
             Academia Sinica, Institute of Astronomy and Astrophysics, Taiwan\\
             }

   \date{Received {\today}; accepted {\today}}

\abstract
  {We present galaxy counts at 15 microns using the Japanese AKARI satelitte's NEP-deep and NEP-wide legacy surveys at the North Ecliptic Pole.  The total number of sources detected are approximately 6700 and 10,700 down to limiting fluxes of 117 and 250 microJy (5 sigma) for the NEP-deep and NEP-wide survey respectively. We construct the Euclidean normalized differential source counts  for both data sets (assuming 80 percent completeness levels of 200 and 270 microJy respectively) to produce the widest and deepest contiguous survey at 15 microns to date covering the entire flux range from the deepest to shallowest  surveys made with the Infrared Space Observatory (ISO)  over areas sufficiently significant to overcome cosmic variance, detecting six times as many sources as the largest survey carried out with ISO. We compare the results from AKARI with the previous surveys with ISO at the same wavelength and the Spitzer observations at 16 microns using the peek-up camera on its IRS instrument. The AKARI source counts are consistent with other results to date reproducing the steep evolutionary rise at fluxes less than a milliJansky and super-Euclidean slopes. We find the the AKARI source counts show a slight excess at fluxes fainter than 200 microJanskys which is not predicted by previous source count models at 15 microns. However, we caution that at this level we may be suffering from the effects of source confusion in our data. At brighter fluxes greater than a milliJansky, the NEP-wide survey source counts agree with the Northern ISO-ELAIS field results, resolving the discrepancy of the bright end calibration in the ISO 15 micron source counts.}

  \maketitle
%

\section{Introduction}\label{sec:introduction}
 
Tiered cosmological surveys at infrared wavelengths are the most efficient method to collect data on large ensembles of dusty star-forming galaxies viewed and different cosmological epochs (i.e. different redshifts). Since the pioneering survey of {\it IRAS} (\cite{soifer87}) more than two decades ago, a series of dedicated infrared missions have pushed our understanding of the dusty Universe to increasingly further cosmological distances. Although,  {\it IRAS}  discovered indications of a dusty evolving Universe, it was the deep mid-infrared surveys carried out with the Infrared Space Observatory ({\it ISO}, \cite{kessler96}), especially in the ISOCAM 15$\mu$m band (e.g.,  \cite{elbaz99}, \cite{elbaz05} and references therein), which revealed the true extent of the violent evolution underway in galaxies from redshifts 0.1-1. This evolution in the infrared galaxy population was confirmed by the later {\it Spitzer} space observatory surveys at 24$\mu$m (\cite{papovich04}) which also extended the observations of the evolution in the infrared galaxy population out to even higher redshifts of $\sim$2.5, thus linking the local Universe to the distant high redshift (z$>$2.5) Universe observed at submillimetre wavelengths by the Submillimetre Common User Bolometer Array (SCUBA, e.g.  \cite{smail97}, \cite{hugh98}, \cite{blain99}, \cite{scott02}, \cite{mortier05})

The {\it AKARI} satellite was launched on board JAXA's M-V8 launch vehicle on February 22, 2006 Japan standard time, JST and is Japan's first space mission dedicated to infrared astrophysics  (\cite{murakami07}). The 68.5 cm cooled telescope directs light on to two  focal plane instruments, the Far-Infrared Surveyor (FIS) (\cite{kawada07}) and the Infrared Camera (IRC) (\cite{onaka07}).  {\it AKARI}'s orbit is Sun-synchronous which dictates that any large surveys can only be carried out at the ecliptic poles. The North Ecliptic Pole (NEP) was chosen as the site for  {\it AKARI}'s premier deep cosmological survey with the IRC instrument (The {\it AKARI}-NEP survey \cite{matsuhara06}). The IRC consists of three cameras, the IRC-NIR, MIR-S \& MIR-L  with nine broad band photometric filters which cover the continuous wavelength from 2 to 26 $\mu$m, in the filter bands N2(2.4$\mu$m),
N3(3.2$\mu$m), N4(4.1$\mu$m), S7(7.0$\mu$m), S9W(9.0$\mu$m), S11(11.0$\mu$m), L15(15.0$\mu$m),
L18W(18.0$\mu$m), and L24(24.0$\mu$m) respectively. The NEP survey has a "wedding cake" configuration with a central  $\sim$ 0.38 square degree deep circular area with $\sim$2500 sec exposures for each filter (The NEP Deep survey \cite{wada08}), and a near concentric surrounding shallower  survey covering 5.8 square degrees with $\sim$300 sec exposures for each filter hereafter referred to as the NEP-wide survey \cite{lee08}. 
\
The real strength of the {\it AKARI} NEP survey lies in the unprecedented photometric coverage from near to
mid-infrared wavelengths, critically including the wavelength domain between {\it Spitzer}'s IRAC and MIPS instruments from 8$< \lambda <$24$\mu$m where only limited coverage is available (from the peek-up camera on the IRS, \cite{teplitz05a}). This multi-wavelength view provides a unique opportunity to
study infrared luminous galaxies whose mid-infrared spectra are complicated by emission and absorption features i.e., 
the PAH lines and silicate absorption feature, etc.

In this work, we report on the  IRC L15 15 $\mu$m band results in the {\it AKARI} NEP survey region. In Section \ref{sec:observations} we summarize the observations and processing of the L15 band data.  Source extraction, photometry and construction of the source counts is presented in Section \ref{sec:Source Counts}. In Section \ref{sec:analysis} we compare the results of the {\it AKARI} L15 band survey with the previous surveys carried out by the {\it ISO} \& {\it Spitzer} space observatories. A brief summary is given in Section  \ref{sec:summary}. Throughout this work, a concordance cosmology of  $H_o = 72kms^{-1}Mpc^{-1}, \Omega=0.3, \Lambda=0.7$ is assumed.

 \smallskip

\section{Observations and data reduction}\label{sec:observations}

\subsection{The {\it AKARI} NEP large-area survey at 15$\mu$m}\label{sec:nepsurvey}

The  NEP survey was carried out as  {\it AKARI} guaranteed time core observations referred to as a large-area survey (LS) program. The stipulation for the NEP survey was to cover a sufficient area to a sufficient depth to overcome the effect of cosmic variance on the resulting statistics (See \cite{somerville04}  for discussion on cosmic variance and  \cite{matsuhara06} for the NEP survey design).  The volume required to sufficiently overcome cosmic variance (achieve a variance $<$5-10$\%$) transpires to areas of $>$0.5 square degrees for redshifts of $>$1 (the NEP-deep survey) and $>$5 square degrees for redshifts $<$1  (the NEP-wide survey). The observations were carried out throughout the mission lifetime over the period of May 2006 to August 2007.
For the NEP-deep survey a total of 266 pointed  observations  were taken tracing out a central circular region with uniform coverage of $\sim$0.38 square degrees centred on a position  RA=17h56m, Dec.=66d37m offset from the NEP, using the  IRC astronomical observing template (AOT) designed for deep observations (IRC05). For each waveband a total of 4 pointings were made at each position. The details of the observation strategy for the NEP-deep survey are given in  \cite{wada08}. Note that the total area covered in the 15$\mu$m band including all coverage around the periphery of the central region results in a total area of $\sim$0.6 square degrees.

The NEP-wide survey consisted of a total of 446 pontings were made in a circular region covering 5.8 square degrees centred on the NEP (i.e. encompassing but slightly offset from the NEP-deep survey). To maximize the area and the number of wavebands, the shallower IRC03 observing template was used. A total of 3 pointings were made at each position. The details of the observation strategy for the NEP-wide survey are given in  \cite{lee08}.

\smallskip
\subsection{Data reduction}\label{sec:Reduction}

The data for the NEP survey were reduced by the standard IRAF-based IRC imaging pipeline (version  20071017; see the IRC data user manual, \cite{lorente07}). The pipeline corrects basic instrumental effects, performing dark subtraction,
linearity correction, distortion correction and flat fielding for individual data frames. Removal of the diffuse background
was achieved by subtracting the median filtered self-image 
from the observations.

Spurious events, such as cosmic rays were removed at the frame co-addition stage using a 3 sigma clipping
technique.  The average value for each pixel among the frames rather than the median  value, was used in order to further improve the signal-to-noise of the final image.
Individual frames are co-add together using associations with bright stars between frames to produce the final pointing image.
 Astrometry was added to the 15$\mu$m image frames  by comparing pointings with those taken at the the shorter, near-infrared  {\it AKARI} wavebands (The near-infrared   {\it AKARI} wavebands have astrometry added by comparison with the 2MASS catalogue).

The individually processed pointings were then mosaiced together using the 
the publicly available software SWarp\footnote{http://terapix.iap.fr/rubrique.php?id rubrique=49} to produce the final mosaiced image. A detailed account of the data reduction is given in \cite{wada08}.

\smallskip

\section{Source counts}\label{sec:Source Counts}

\subsection{Source extraction and photometry}\label{sec:extraction}

\begin{figure}
\centering
\centerline{
\psfig{figure=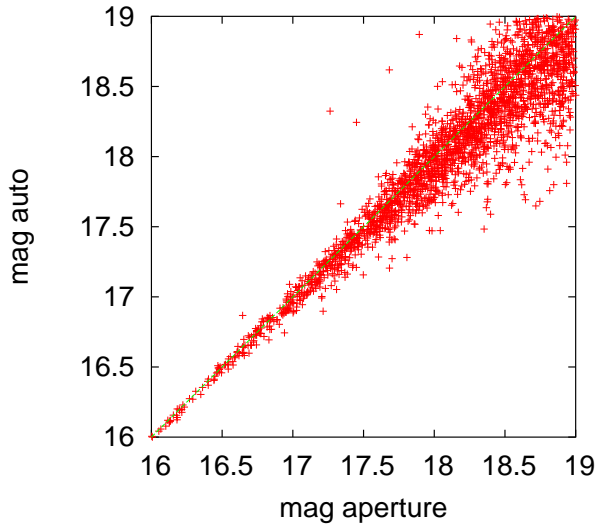,width=8cm}
}
\caption{The result of the source extraction using the MAGAPER and MAGAUTO parameters in SExtractor. The MAGAPER assumes apertures equal to the same aperture radius of standard stars used for the flux calibration while the MAGAUTO fits the best elliptical aperture. 
\label{aperturecorrection}}
\end{figure}

Sources were extracted from the final NEP-deep and NEP-wide images using the  SExtractor software (\cite{bertin96}) with the criteria of five connected pixels in the image map having more than a 1.65$\sigma$ signal above the local background, corresponding to a  5$\sigma$ detection for a uniformly distributed flux. 
Photometry was carried out using SExtractor's MAGAUTO variable elliptical aperture size technique with the aperture parameters ÒKron factorÓ and Òminimum radiusÓ set to their default values of 2.5 and 3.5,
respectively and with the magnitude zero point derived from observations of standard
stars (\cite{tanabe08}). The source extraction parameters are summarized in Table \ref{tab:parameters}. Aperture corrections were estimated by carrying out fixed aperture size photometry at different apertures from 5.0,6.0,9.0,12.0,15.0,20.0 arcsecs using  SExtractor's MAGAPER function. In Figure \ref{aperturecorrection} we show the result of the source extraction to the the 5$\sigma$ level of the NEP survey using the MAGAPER and MAGAUTO parameters in SExtractor. The MAGAPER assumes apertures equal to the same aperture radius of standard stars used for the flux calibration while the MAGAUTO fits the best elliptical aperture. It was found that with the aperture set to the same size  as used for the standard star observations (7.5 pixels for the L15 band image), that both aperture techniques were consistent with each other and that no aperture correction was necessary. 

\begin{table}
\caption{ Summary of parameters used for source extraction}
\centering
\begin{tabular}{@{}ll}
\hline\hline
Parameter& Value    \\
\hline
Pixel Scale & 2.34 arcsec  \\
Detection Threshold & 1.65  \\
Number of pixels required above threshold & 5  \\
Aperture diameter & 7.5 arcsec  \\
Kron factor  & 2.5 arcsec \\
Minimum radius  & 3.5 arcsec \\
magnitude zero point & 23.573\\
\hline
\end{tabular}
\label{tab:parameters}
\end{table}

\begin{figure}
\centering
\centerline{
\psfig{figure=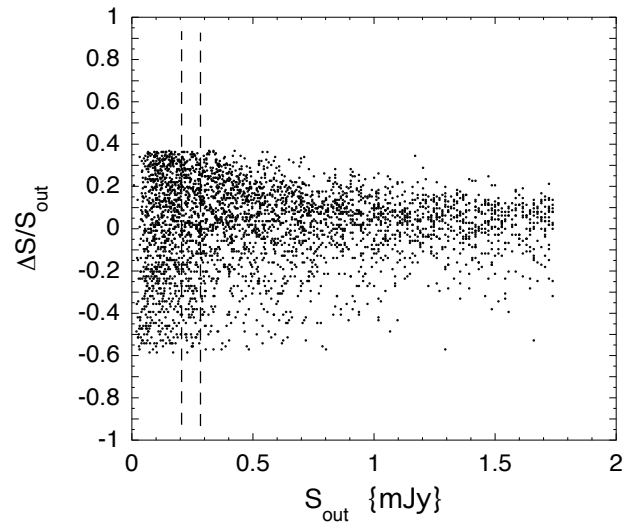,width=8cm}
}
\caption{A simulation of the photometric accuracy of our source extraction. The fractional difference in the flux $\Delta$S/S$_{out}$ ([output - input]/output) as a function of the measured flux density. Also shown are the 80$\%$ completeness limits from Section \ref{sec:completeness}. 
\label{photometricaccuracy}}
\end{figure}  

\begin{figure}
\centering
\centerline{
\psfig{figure=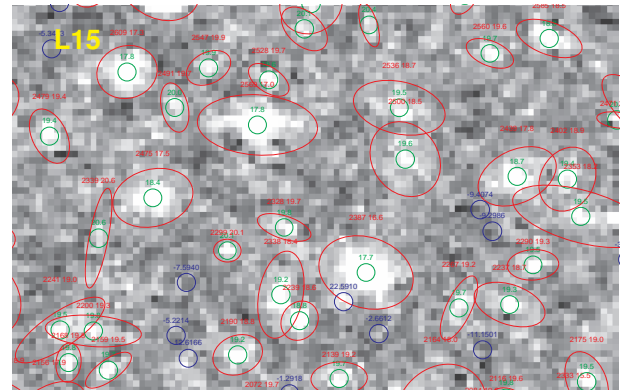,width=8cm}
}
\caption{A segment of the {\it AKARI} NEP Deep survey field showing the typical distribution of sources. The small circles mark the position of the sources where the peak flux is measured. The larger surrounding ellipses are the MAGAUTO apertures. The smaller isolated circles not on a source are areas of blank sky measurements.
\label{blending}}
\end{figure}  

\begin{figure}
\centering
\centerline{
\psfig{figure=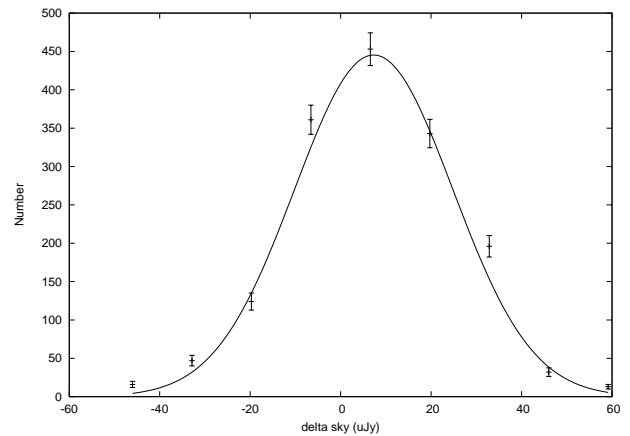,width=8cm}
}
\caption{The noise histogram for the {\it AKARI} NEP survey. The measured points are well fitted by a Gaussian with a tail to positive fluxes indicative of a faint source population (contributing to the confusion).
\label{noisehistogram}}
\end{figure}  

In Figure  \ref{photometricaccuracy} we create a simulation to  check the  photometric accuracy of our source extraction. Artificial sources are input to the image and then extracted using the same source extraction procedure as used for the real data. The value $\Delta$S/S$_{out}$ is plotted as a function of the measured flux density where $\Delta$S is the difference between the recovered flux and input flux. We find a dispersion of 15-30$\%$ is seen from bright to fainter fluxes. This values are consistent with the expected flux accuracy of the IRC processing pipeline \cite{lorente07}. Also shown in the figure are the 80$\%$ completeness limits from Section \ref{sec:completeness}.

Figure  \ref{blending} shows a segment of the {\it AKARI} NEP Deep survey field showing the typical distribution of sources. We find that in almost all cases that the sources are sufficiently separated as to avoid any significantly effects from blending.
Calibration of the source fluxes was made using the conversion factors given in the  the IRC data user manual, (\cite{lorente07}). The final source catalogues consist of  a total of 6737 \& 10686  sources detected at 15$\mu$m in the NEP-deep and NEP-wide surveys respectively.
The noise was estimated  by measuring the fluctuations at random blank sky positions (See Figure  \ref{blending}). Positions close to sources were avoided to ensure no contamination from the sources themselves. We made simple aperture photometry at each random position using the IRAF/PHOT package. The size of aperture radius was set to 1.5 pixels with no weighting or PSF fitting assumed. In order to evaluate the fluctuations in the photometry, the noise histogram  was fit with a Gaussian distribution. We defined the one sigma level of the fluctuation as the standard deviation. The noise histogram is shown in Figure  \ref{noisehistogram}.  Note the positive tail corresponding to faint sources.  

\smallskip

\smallskip

\subsection{Raw source counts}\label{sec:raw}

\begin{figure*}
\centering
\centerline{
\psfig{ figure= 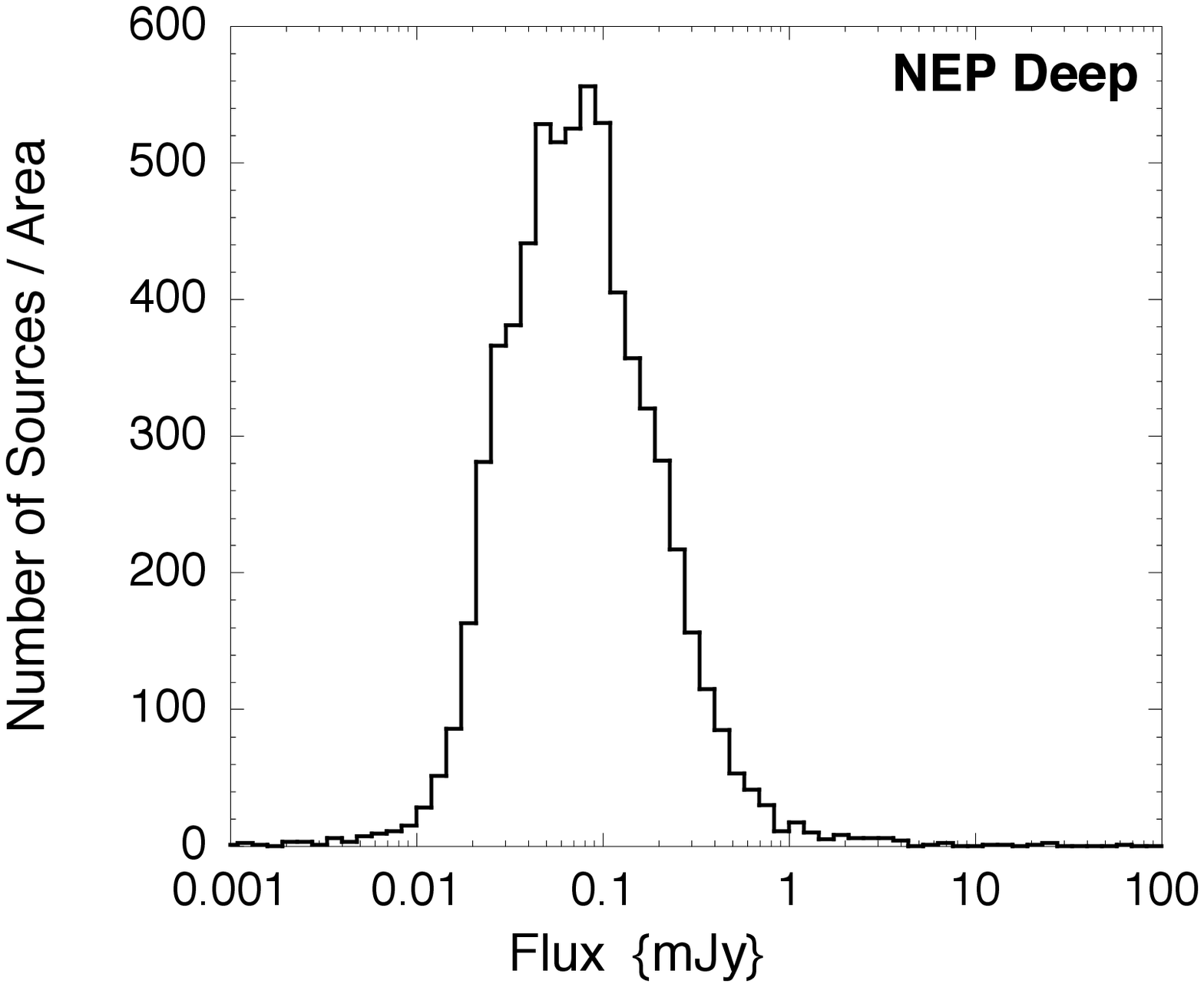,width=9cm}
\psfig{ figure= 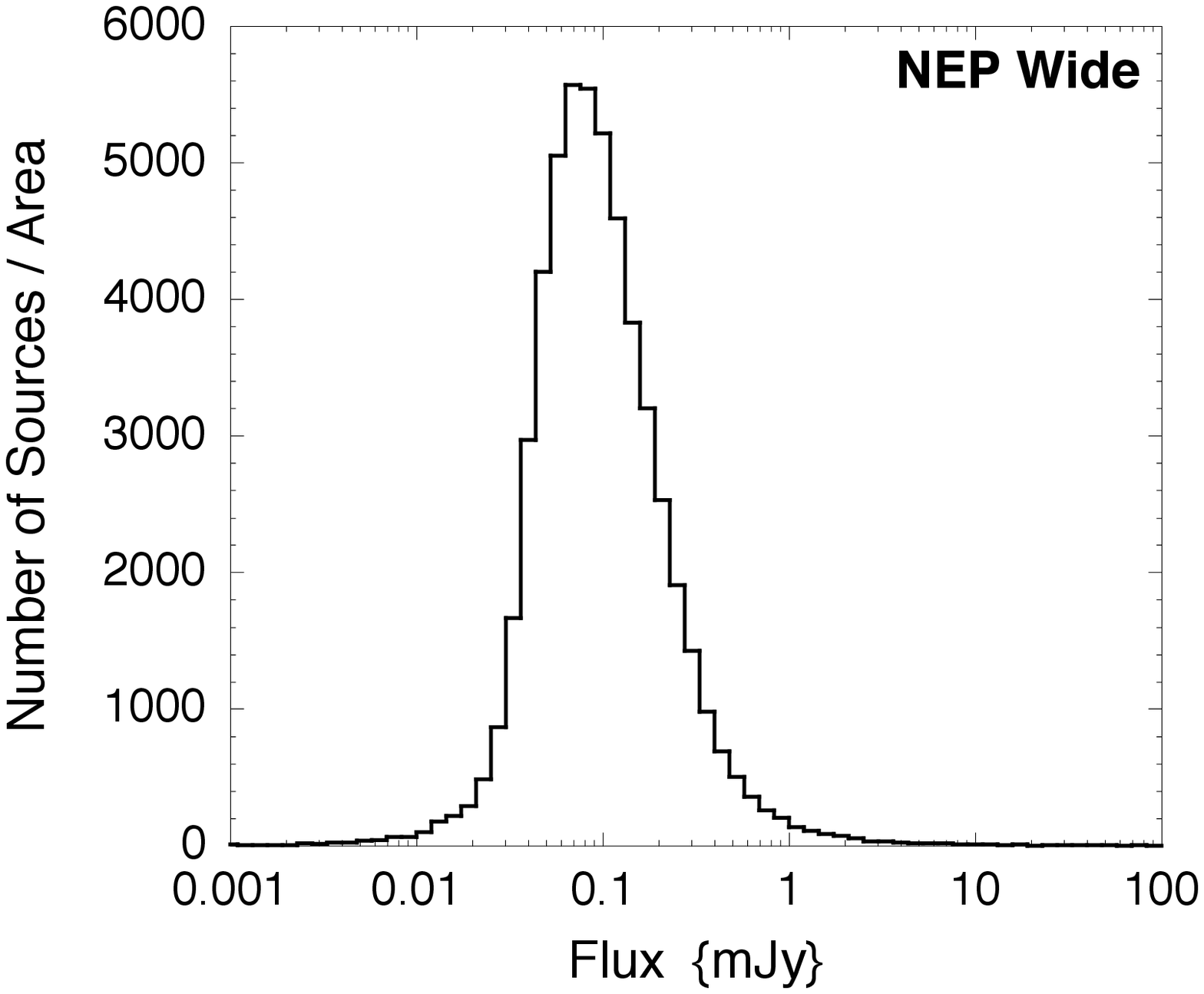,width=9cm}
}
\caption{Raw source counts for the {\it left} {\it AKARI} NEP Deep survey and  {\it right} NEP Wide survey in the L15 (15$\mu$m) band. Differential histogram of the number of sources detected in a flux bin of size $\Delta lgS$=0.08.
\label{rawhistogram}}
\end{figure*}

To calculate the raw source counts from of catalogue we bin the data in flux bins of $\Delta lgS=0.08$. The resulting raw count histogram is shown in the {\it left} panel of Figure \ref{rawhistogram} for the NEP-deep survey and the  {\it right} panel for the NEP-wide survey. For the NEP-deep survey, a total of 6737 sources are detected with a  5$\sigma$ flux limit of $\sim$117$\mu$Jy  (i.e. detected as  5$\sigma$ sources above the noise), although sources are extracted to an order of magintude deeper. We observe a peak in the raw counts histogram between 50-100$\mu$Jy. For the NEP-wide survey a total of 10686 sources are detected at a 5$\sigma$ limit of $\sim$250$\mu$Jy (although the raw numbers were extracted to an order of magnitude deeper). The peak of the distribution, around a milli-Jansky, is somewhat narrower than the NEP-deep survey

\smallskip

\smallskip

\subsection{Completeness, reliability and stellar fractions}\label{sec:completeness}

\begin{figure*}
\centering
\centerline{
\psfig{ figure=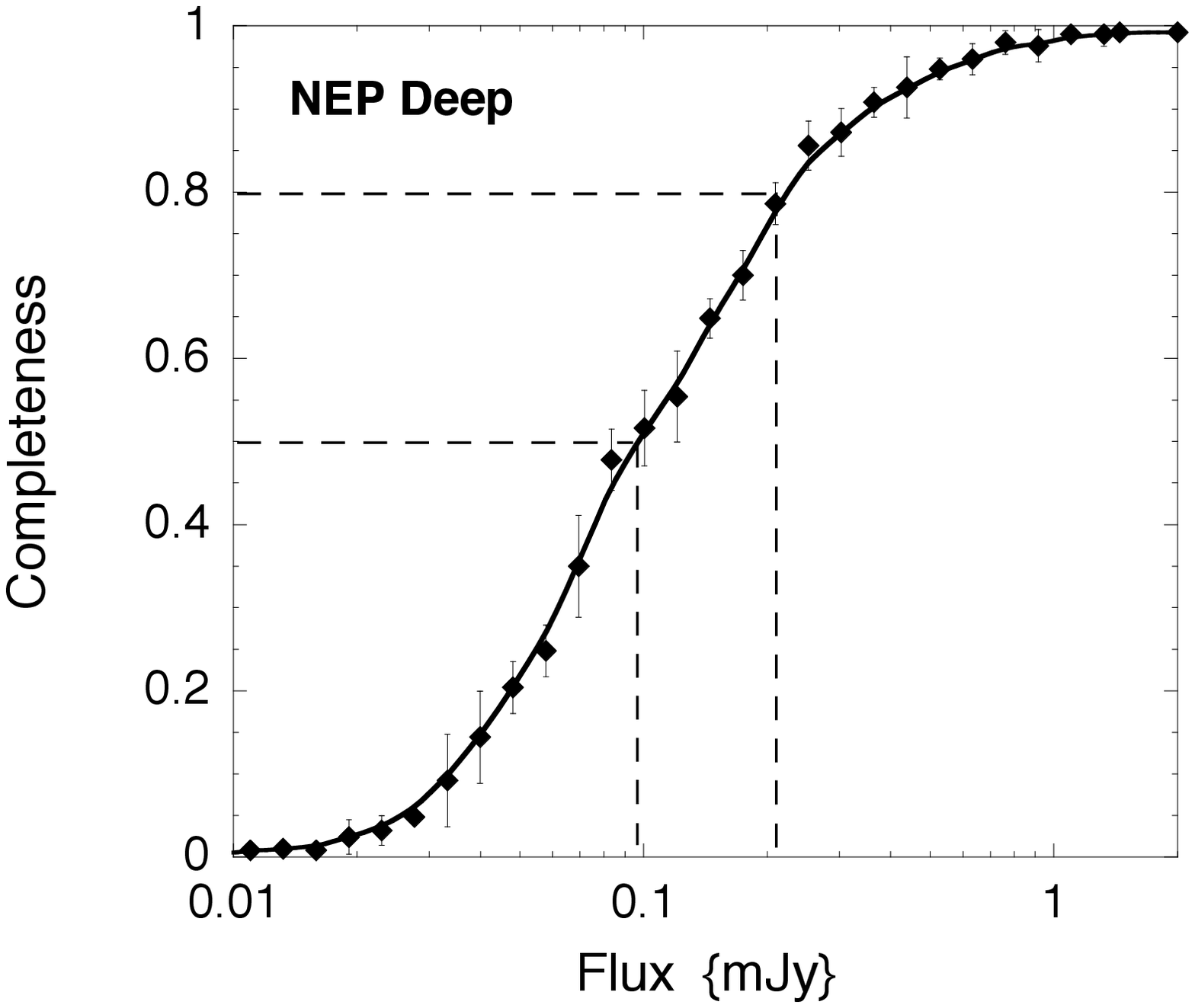,width=8cm}
\psfig{ figure=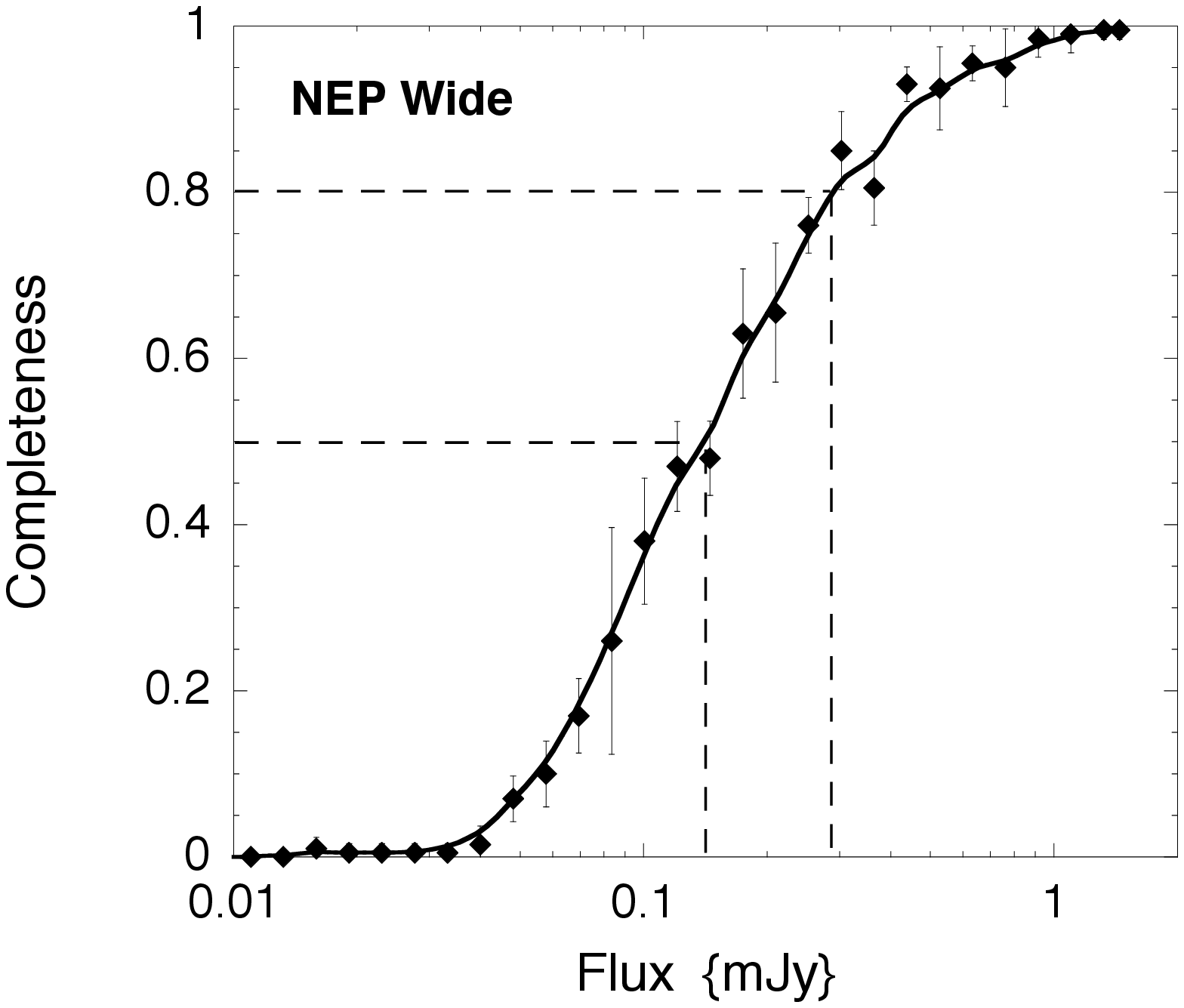,width=8cm}
}
\caption{5$\sigma$ completeness function  for the {\it AKARI} NEP survey in the L15 (15$\mu$m) band for the {\it left} {\it AKARI} NEP Deep survey and  {\it right} NEP Wide survey. The best fit curve to the data and the 50$\%$ and 80$\%$ completeness levels are also shown.
\label{completeness}}
\end{figure*}

The method used for the correction of any incompleteness in the data has been described in detail in \cite{wada08}. The completeness was calculated from Monte Carlo simulations by injecting artificial sources into the map at known positions and then performing the same source extraction to recover the sources. The flux range was divided into logarithimic flux bins of equal size and populated with 20 sources assuming a Euclidean Universe distribution and a circular PSF. The completeness fraction as a function of flux is then given by the ratio of the number of true sources extracted to the total number of input sources. For each bin, five independent simulations were made resulting in around 100 sources per flux bin.
The final  completeness for the survey is shown in Figure \ref{completeness} for the NEP-deep survey ({\it left panel}) and NEP-wide survey  ({\it right panel}) respectively. From the figures, we find 50$\%$ completeness limits of $\sim$100$\mu$Jy and 150$\mu$Jy and 80$\%$ completeness limits of $\sim$200$\mu$Jy and 270$\mu$Jy for the NEP-deep and NEP-wide surveys respectively. 

\begin{figure*}
\centering
\centerline{
\psfig{ figure=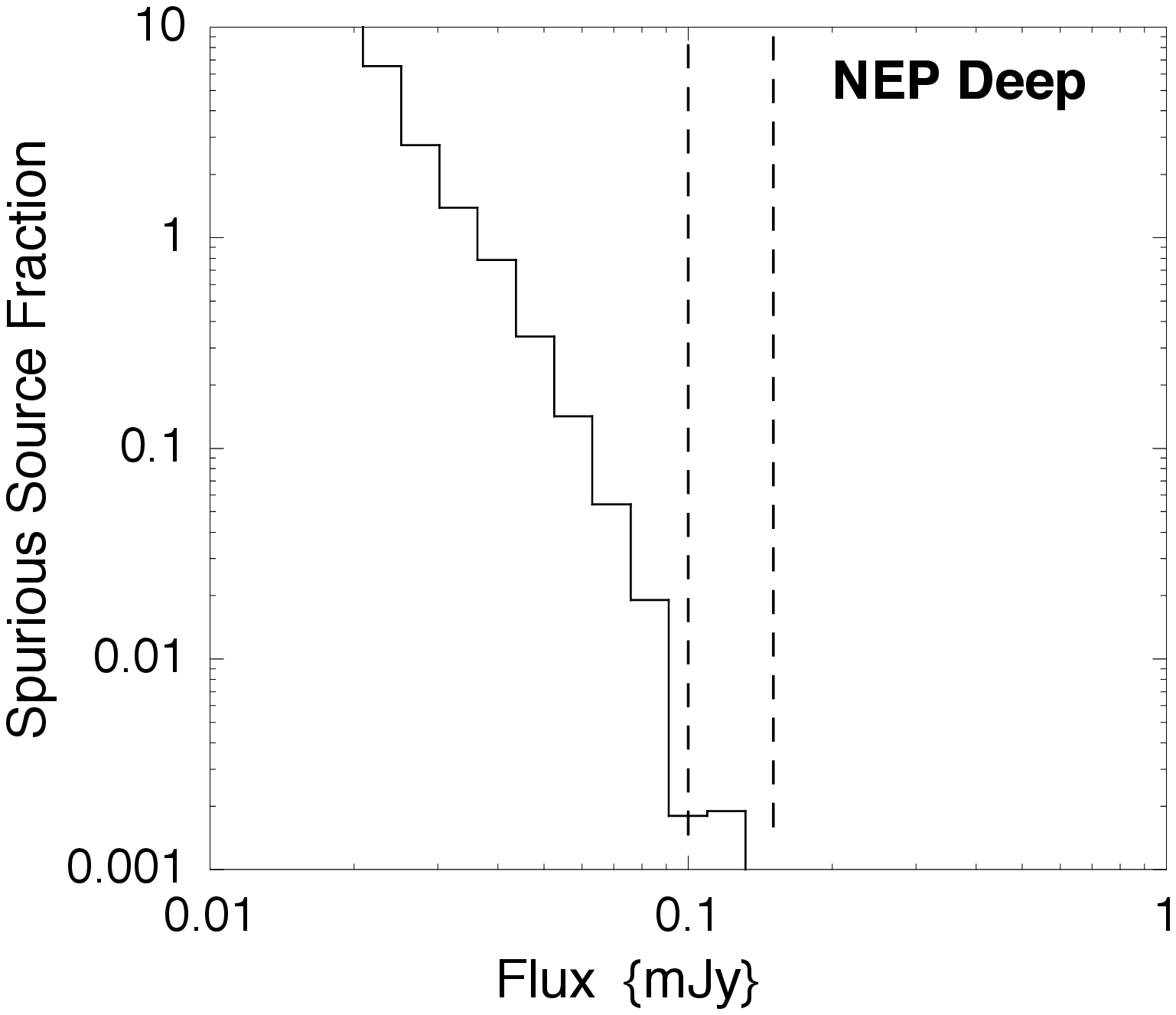,width=8cm}
\psfig{ figure=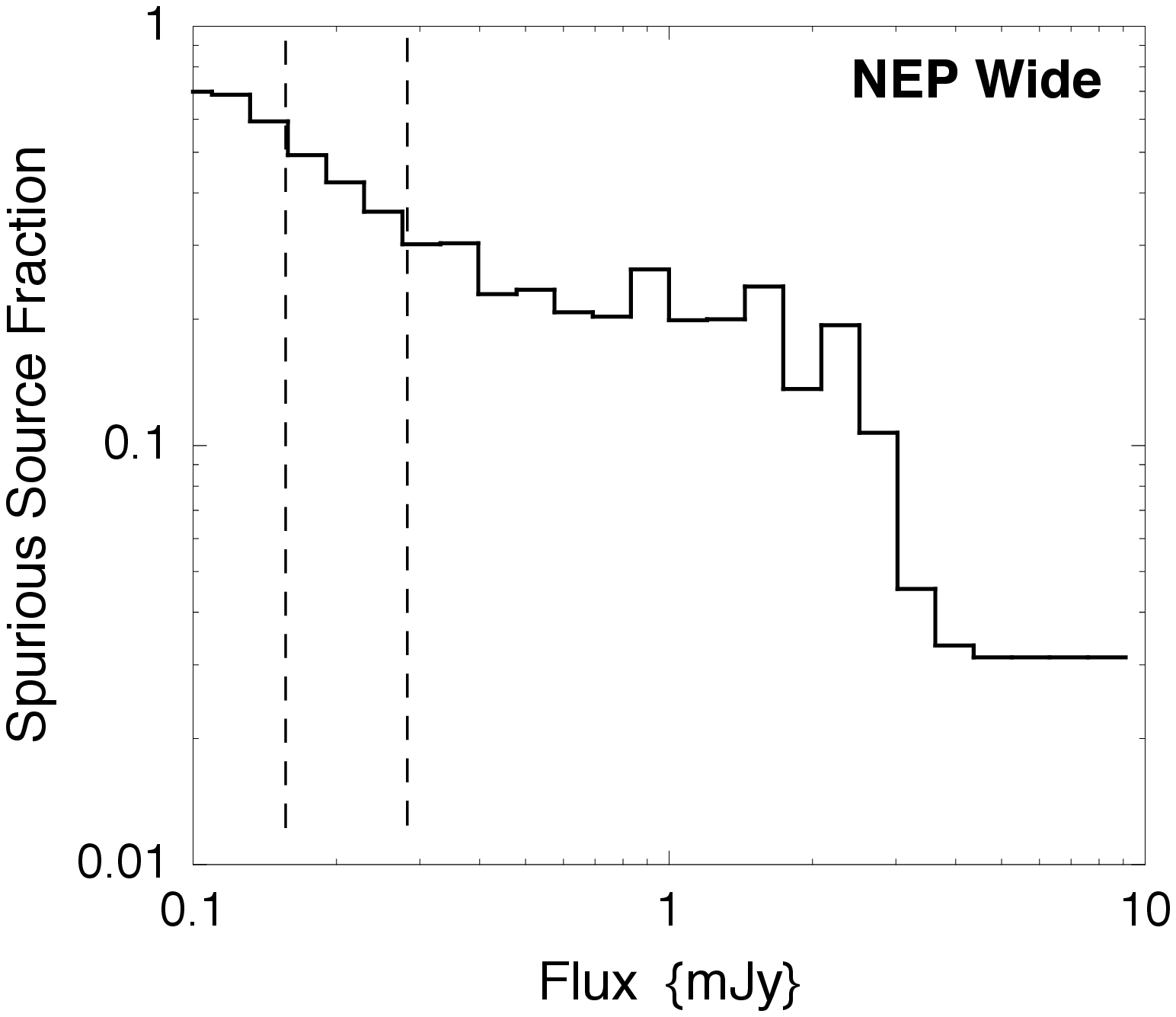,width=8cm}
}
\caption{A measure of the reliability of the NEP surveys measured by examining the fraction of spurious sources detected  from the negative image of the the {\it left} {\it AKARI} NEP Deep survey and  {\it right} NEP Wide survey. The 50$\%$ and 80$\%$ completeness levels are also shown for comparison.
\label{reliability}}
\end{figure*}

The reliability of the detections has been checked by carrying out the same source extraction on the negative images and calculating the fraction of these spurious sources detected to the number of real sources detected on the positive image as a function of flux. The results are shown in Figure \ref{reliability} for the fraction of spurious detections. We find that the NEP-deep survey is expected to be highly reliable at both the 50$\%$ and the 80$\%$ completeness level. However, in the case of the NEP-wide survey, the survey is only  50$\%$ reliable (following this definition of reliability) at the  50$\%$ completeness level. At the  80$\%$  completeness level the NEP-wide survey is $>$75$\%$ reliable. As an additional reliability check, the overlapping regions of the NEP-deep and -wide surveys were also examined by eye to confirm the reality of sources in both images.  We find that the NEP-wide sources can be reliably identified with NEP-deep counterparts down to a  flux limit of $\sim$250$\mu$Jy which is brighter than the measured 50$\%$ completeness limit of 150$\mu$Jy discussed above. Therefore we assume  the  80$\%$ completeness limit numbers  for our source count analysis.   

Stellar fractions were calculated using the optical data taken by the Canada France Hawaii Telescope (CFHT) and Near-Infrared data taken by the Kitt Peak National Observatory (KPNO) over the NEP-wide region (which also encompasses the NEP Deep region (\cite{hwang07})). The  {\it AKARI} 15$\mu$m sources were cross matched with CFHT/KPNO data, with stellar sources have stellarity $>$ 0.8 and optical $r'$ band magnitudes $<$ 19 identified during the source extraction. The stellarity was checked via colour-magnitude and colour-colour diagrams. It is found that almost all the star-like sources have bright  {\it AKARI} L15 band AB magnitudes $<$14 while the extragalactic sources lie at magnitudes $>$14. The stellar sources can also be clearly segregated using the stellarity $>$ 0.8 and optical $r'$band magnitudes $<$ 19 and the near-infrared colour criteria of H-N2$<$-1.6 in the  H-N2, g-H colour plane (where N2 is the  {\it AKARI} IRC 2.5$\mu$m band). We have also checked our stellar criteria in the H-L15 versus H Color-Magnitude diagram (similar to Shupe et al. 2008, but with somewhat different filter bands). Our criteria agree well (within a few $\%$) and therefore is regarded as sufficient in statistically correcting for stellar sources. In Table \ref{tab:stellarcontribution} the stellar fraction as a function of flux (bins of $lg(S/mJy)$=0.2) and the star counts are also shown in Figure  \ref{dctcounts}. 

\begin{table}
\caption{ Fraction of stellar contribution to source counts}
\centering
\begin{tabular}{@{}ll}
\hline\hline
lg(Flux) & Stellar fraction    \\
(mJy) &    \\
\hline
1.272 & 0.5   \\
1.072 & 0.35   \\
0.872 & 0.286   \\
0.672 & 0.20   \\
0.472 & 0.136   \\
0.272 & 0.135   \\
0.072 & 0.133   \\
-0.128 & 0.093   \\
-0.328 & 0.091   \\
-0.528 & 0.082   \\
-0.728 & 0.063   \\
-0.928 & 0.054   \\
\hline
\end{tabular}
\label{tab:stellarcontribution}
\end{table}

\smallskip

\subsection{Confusion and clustering}\label{sec:confusion}
At the faint fluxes probed by the NEP-deep survey, confusion - defined as the flux limit below which sources are indistinguishable from the background noise becomes an important limiting factor. Confusion noise is a function of the beam size of the telescope and is thus worse at longer wavelengths and smaller apertures. Classically, the confusion limit is defined at the level where one source occupies a beam but practically the confusion limit is often set to 20-50 beams per source (\cite{hogg01}, \cite{condon74}). For  {\it ISO} \cite{oliver97} carried out a $P(D)$ analysis of the  {\it ISO}-HDF observations which did not use a source detection algorithm but rather looked for fluctuations in the background caused by sources around or below the confusion limit in the survey. A clear, positive signal was indeed detected in the data and interpreted as the contribution from faint galaxies.  \cite{oliver97} concluded that for  {\it ISO}, a 60cm diameter telescope, the confusion limit due to point sources was $\sim$200$\mu$Jy. For  {\it AKARI}, a 70cm diameter telescope, the classical confusion limit of 50 beams per source is  predicted to be  $\sim$100$\mu$Jy using the models of \cite{pearson05}. However at the faint sensitivities of the NEP-deep survey, where the source counts are strongly non-Euclidean the classical beams per source criteria may not be accurate. To investigate the confusion limit of the NEP survey we carried out a measurement of the beam to beam fluctuations in the noise as a function of flux (e.g. \cite{jeong06}, \cite{dole03}, \cite{vaisanen01}) by defining the signal to noise, $q=S_{lim}/\sigma_{c}(S_{lim})$, as the ratio of the flux limit to the rms noise measured from beam to beam fluctuations due to sources fainter than the flux limit. The equation is solved iteratively with values for  $q$ usually chosen to be between 3-5.  We find values of $S_{lim}$=25 - 53$\mu$Jy for $q$=3 \& 5 respectively. \cite{wada08} have calculated the number of sources per beam for the NEP-deep survey finding that for the L15 band there is an average of 1/53.7 sources per beam at the 5$\sigma$ detection limit of 117$\mu$Jy. This value of 53.7 beams per source is close to the classical confusion limits quoted by  (\cite{hogg01}, \cite{condon74})

We have also used the a source density criterion as detailed by,  \cite{jeong06}, \cite{dole03} for the case where the fluctuations are not the dominant contribution to the confusion but rather the the high density of resolved sources around the sensitivity limit (as is the case in the mid-infrared, e.g. \cite{dole04}, \cite{rodighiero06}).  In order to find the flux limit where the source detection becomes affected by confusion due to near neighbours we set the constraint of 80$\%$ completeness and use simulations to calculate $S_{lim}$  and calculate the corresponding fluctuations at that level. The simulations yield confusion limits of $\sim$130$\mu$Jy higher than the values from the fluctuation criteria alone but consistent with the  53.7 beams per source value of  \cite{wada08} and closer to the theoretical prediction using the models of  \cite{pearson05}.

Note that the  {\it AKARI} NEP surveys are specifically designed to overcome major effects of cosmic variance including the effect of clustering on the counts (See Section \ref{sec:nepsurvey}). Although an analysis of the clustering properties of the NEP field is beyond the scope of this paper and will be presented in later work, we note that the amplitude due to clustering in the {\it Spitzer} 24$\mu$m surveys has been found to be of the order of 0.0009-0.01 for $S_{24}>$350$\mu$Jy  (\cite{magliocchetti07}, \cite{magliocchetti08}), where the amplitude $A$ is related to the angular correlation function $w(\theta )= A \theta^{1-\gamma}$, where $\theta$ is the angular separation on the sky and  $\gamma$ is the clustering index.  \cite{takeuchi01} have shown that the errors on the number counts due to clustering depend on the angular correlation function and the area of the survey and provide a formulation for the signal to noise given by $S/N=\sqrt{\Omega / \int w(\theta) d\Omega}$, where $\Omega$ is the solid angle of the survey. Using this formulation we find errors on the number counts due to clustering between 3-9$\%$ for redshifts $<$1.6. 

\smallskip

\subsection{Final L15 band source counts}\label{sec:counts}

\begin{figure*}
\centering
\centerline{
\psfig{ figure=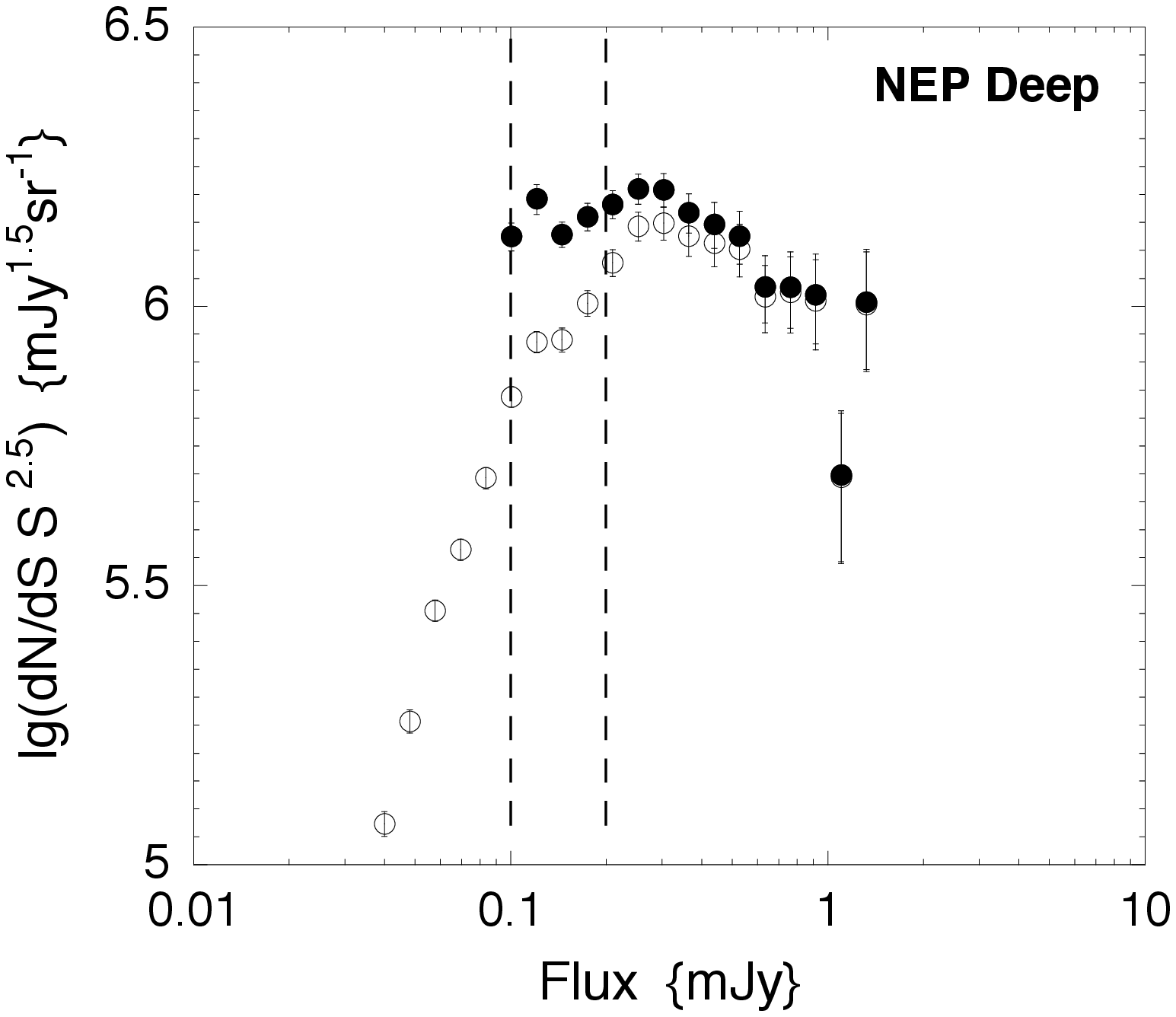,width=9cm}
\psfig{ figure=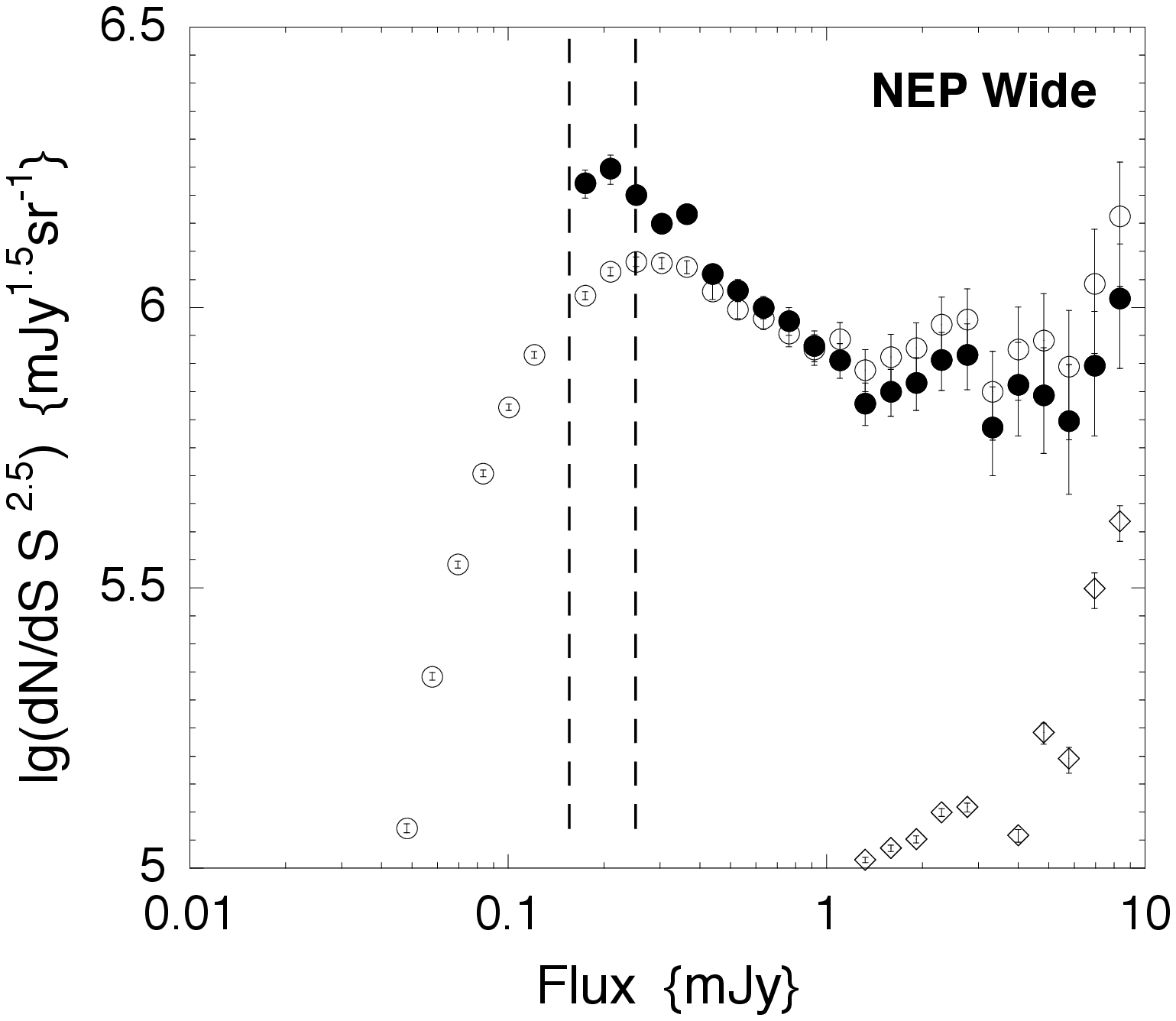,width=9cm}
}
\caption{Completeness corrected differential source counts ({\it filled circles}) normalized for a Euclidean Universe  in the {\it AKARI} L15 band for the  {\it left} NEP Deep survey and  {\it right} NEP Wide survey. The vertical dashed lines show the 50$\%$ and 80$\%$ completeness levels from Figure \ref{completeness} . The raw data is shown as {\it unfilled circles}. Also shown for the NEP Wide survey are the stellar counts ({\it unfilled diamonds}).
\label{dctcounts}}
\end{figure*}

Using the results from Section \ref{sec:completeness} we present the completeness and stellar corrected normalized differential source counts per steradian ($(dN/dS) S^{2.5}$ in the units of mJy$^{1.5} $) in the {\it AKARI} L15 15$\mu$m band in Figure \ref{dctcounts} for the NEP-deep survey ({\it left panel}) and NEP-wide survey  ({\it right panel}) respectively. The Euclidean normalized differential sources are characterized by a flat distribution at bright fluxes, since the counts are normalized to a flux, S$^{2.5}$. Any evolution will present itself as an upturn in the counts at fainter fluxes and the counts will eventually tail off to fainter fluxes as cosmological  redshift dimming takes effect.  The counts are plotted for fluxes brighter than the 50$\%$ completeness limit of  $\sim$100$\mu$Jy \& 150$\sim$100$\mu$Jy for the NEP-deep and  NEP-WIDE surveys respectively although the counts are assumed to be actually reliable at the 80$\%$ completeness limits of  $\sim$200$\mu$Jy \& 270$\sim$100$\mu$Jy  We also show the original raw source counts corrected to be counts per steradian and the star counts in the NEP-wide plot for comparison. At flux densities 1 $<$ S $<$10 mJy the source counts are Euclidean in nature as expected. We see the characteristic evolutionary rise at fluxes below 1mJy producing super-Euclidean slopes as seen in the {\it ISO} 15$\mu$m surveys (\cite{fadda04}, \cite{metcalfe03}, \cite{gruppioni02},  \cite{altieri99}, \cite{aussel99}). The differential source counts in the  {\it AKARI} L15  band begin to rise from the Euclidean case around a flux of a mJy with the peak occuring between flux densities 0.25 $<$ S $<$0.3 mJy. At fainter flux densities ($<0.2mJy$) the source counts begin to fall away.
The completeness corrected normalized differential source counts, $\mathrm{d}N/\mathrm{d}S . S^{2.5}$, from Figure  \ref{dctcounts},  are tabulated in Tables \ref{tab:deepcounts} \&  \ref{tab:widecounts} for the NEP-deep and NEP-wide surveys respectively for the 50$\%$ completeness limit.
Note that by summing the differential counts  to obtain the integral number counts, we find that the NEP-wide source counts brighter than a milli-Jansky exhibit a slope of around 1.5, i.e. consistent with the non-evolving Euclidean Universe while at fainter fluxes both the NEP-wide and -deep counts are above Euclidean expectations.

\begin{table}
\caption{ {\it AKARI}  band NEP-deep survey L15 band  Euclidean normalized differential source counts.}
\centering
\begin{tabular}{@{}llll}
\hline\hline
lg(Flux) & Counts& \multicolumn{2}{c}{Errors}   \\
&  $\mathrm{d}N/\mathrm{d}S . S^{2.5}$ &(low) & (high)\\
 (mJy) & (mJy$^{1.5}$sr$^{-1}$) & \multicolumn{2}{c}{(mJy$^{1.5}$sr$^{-1}$)} \\
\hline
0.121839 & 6.007522 & 5.886836 & 6.101860 \\
0.041839 & 5.698483 & 5.542632 & 5.812936 \\
-0.038160 & 6.020391 & 5.932722 & 6.093298 \\
-0.118160 & 6.034276 & 5.960460 & 6.097348 \\
-0.198160 & 6.034723 & 5.970412 & 6.090726 \\
-0.278161 & 6.125328 & 6.075381 & 6.170118 \\
-0.358161 & 6.146805 & 6.103473 & 6.186158 \\
-0.438161 & 6.167757 & 6.131282 & 6.201397 \\
-0.518161 & 6.208664 & 6.177380 & 6.237808 \\
-0.598160 & 6.210495 & 6.182887 & 6.236405 \\
-0.678160 & 6.182443 & 6.156610 & 6.206784 \\
-0.758161 & 6.160284 & 6.135064 & 6.184017 \\
-0.838161 & 6.128609 & 6.105217 & 6.150735 \\
-0.918160 & 6.192664 & 6.164233 & 6.218212 \\
-0.998160 & 6.125141 & 6.098685 & 6.149198 \\
\hline
\end{tabular}
\label{tab:deepcounts}
\end{table}

\begin{table}
\caption{ {\it AKARI}  band NEP-wide survey L15 band Euclidean normalized differential source counts.}
\centering
\begin{tabular}{@{}llll}
\hline\hline
lg(Flux) & Counts& \multicolumn{2}{c}{Errors}   \\
&  $\mathrm{d}N/\mathrm{d}S . S^{2.5}$ &(low) & (high)\\
 (mJy) & (mJy$^{1.5}$sr$^{-1}$) & \multicolumn{2}{c}{(mJy$^{1.5}$sr$^{-1}$)} \\
\hline
0.92184 & 6.1625 & 6.0376 & 6.2594 \\
0.84184 & 6.0425 & 5.9176 & 6.1394 \\
0.76184 & 5.8945 & 5.7637 & 5.9949 \\
0.68184 & 5.9408 & 5.8365 & 6.0249 \\
0.60184 & 5.9256 & 5.8344 & 6.0008 \\
0.52184 & 5.8498 & 5.7632 & 5.9219 \\
0.44184 & 5.9788 & 5.9160 & 6.0336 \\
0.36184 & 5.9697 & 5.9150 & 6.0182 \\
0.28184 & 5.9278 & 5.8781 & 5.9724 \\
0.20184 & 5.9118 & 5.8679 & 5.9518 \\
0.12184 & 5.8909 & 5.8517 & 5.9268 \\
0.041839 & 5.9481 & 5.9161 & 5.9779 \\
-0.038160 & 5.9315 & 5.9030 & 5.9582 \\
-0.11816 & 5.9758 & 5.9501 & 5.9999 \\
-0.19816 & 6.0000 & 5.9797 & 6.0193 \\
-0.27816 & 6.0305 & 6.0103 & 6.0495 \\
-0.35816 & 6.0603 & 6.0454 & 6.0746 \\
-0.43816 & 6.1664 & 6.1502 & 6.1817 \\
-0.51816 & 6.1496 & 6.1345 & 6.1637 \\
-0.59816 & 6.2009 & 6.1884 & 6.2128 \\
-0.67816 & 6.2476 & 6.2193 & 6.2716 \\
-0.75816 & 6.2219 & 6.1947 & 6.2451 \\
-0.83816 & 6.2994 & 6.2790 & 6.3175 \\
\hline
\end{tabular}
\label{tab:widecounts}
\end{table}

\smallskip

\section{Analysis}\label{sec:analysis}

\subsection{Comparison with other surveys}\label{sec:surveys}

The wavelength regime on or around  15$\mu$m has enjoyed copious coverage by both the  {\it ISO} \& {\it Spitzer} observatories. The surveys carried out by {\it ISO}-ISOCAM at 15$\mu$m covered survey areas from $\sim 0.001 - 10$ square degrees to depths of 0.1 - 0.5 mJy.  Although {\it Spitzer} strictly did not have imaging capability between the IRAC 8$\mu$m band and the MIPS 24$\mu$m band, it did have the potential for limited photometry at 16$\mu$m (and 22$\mu$m) using the "Peek Up Imaging" (PUI) mode on the Infrared Spectrograph (IRS) instrument (\cite{houck04}).

Table ~\ref{surveys} lists the major surveys to date carried out with the  {\it AKARI}, {\it ISO} and {\it Spitzer} satellites. The table lists surveys by mission, then by decreasing areal coverage. The characteristic sensitivity and total number of sources  detected is also tabulated. We also give some representation of the co-moving volume encompassed by each survey by assuming the spectrum of the archetypical starburst galaxy M82 at a luminosity of $L_{IR} \sim 10^{11}L_{\sun}$ and using the redshift at which this galaxy could be detected in each survey to calculate the corresponding volume.

The largest 15$\mu$m survey carried out with  {\it ISO} was the ELAIS survey (\cite{mrr04}, \cite{vaccari05}, \cite{pozzi03}). The resulting catalogue consisted of 1546 sources detected with S/N $>$ 5 in the 0.5 -- 100 mJy flux range over a total area of 10.3 square degrees in five fields (N1, N2, N3, S1, S2). ELAIS is best compared with the {\it AKARI} NEP-wide survey covering around half the area but to  a factor of two - three deeper in flux, detecting more than six times the number of sources. The superior depth of NEP-wide should enable the detection of  $L_{IR}>10^{11}L_{\sun}$ luminous infrared galaxies out to a redshift of unity compared to z$\sim$0.4-0.5 for the ELAIS survey.

The {\it AKARI} NEP-deep survey is almost as deep as the very deepest surveys carried out with  {\it ISO} in the lensed cluster fields (\cite{altieri99}, \cite{metcalfe03}). However, the cosmological volume sampled by the NEP-deep survey is thirty-seven times larger resulting in the detection of 1000's rather than 10's of sources, without the constraint of any complicated lens corrections.

Surveys of the Hubble deep fields (HDF), North and South have been carried out by  {\it Spitzer} (and previously by {\it ISO}) using the IRS-PUI at 16$\mu$m.  (\cite{teplitz05a}, \cite{teplitz05b}). These observations, along with the {\it AKARI} PV-phase observations of \cite{wada07} are the deepest 15$\mu$m observations to date  reaching to depths of 40-90$\mu$Jy (but see \cite{hopwood10} for deeper initial results with  {\it AKARI}). However, the areas covered are tiny (10-150 square arcmins) and corresponding volumes ($<$0.2 Mpc$^{3}$) subject to the effects of cosmic variance (\cite{somerville04}).

\subsection{Comparison with observed source counts}\label{sec:models}

In Figure \ref{isospitzercounts} we compare the results from the  {\it AKARI} NEP-deep and -wide surveys with the observed source counts from the various surveys carried out by the  {\it ISO} \& {\it Spitzer} observatories at 15$\mu$m. The figure shows the normalized differential counts per steradian showing that the  {\it AKARI} source counts are in general agreement with the differential source counts at 15$\mu$m to date, reproducing the upturn at around a milli-Jansky and the bump at around the 0.3mJy level. However, the fainter end of the  {\it AKARI}  differential source counts falls off significantly slower than the corresponding {\it Spitzer} observations (Note that the  {\it Spitzer} counts have not been corrected for completness).
The results from the NEP-deep survey exhibit the same excess at fainter fluxes detected in the   {\it AKARI} NEP PV-phase observations by \cite{wada07}.  The  {\it AKARI} counts are more consistent (but still somewhat higher) with the deepest lensed surveys carried out with  {\it ISO} \cite{altieri99}.

At bright fluxes, the  {\it AKARI} NEP-wide source counts provide a new constraint on the bright end source counts. Note that at fluxes of  1-10mJy, the {\it ISO} source counts in the ELAIS-S field \cite{gruppioni02} are systematically lower than the Northern ELAIS fields and the fainter ISO surveys  (\cite{serjeant00}, \cite{vaisanen02}, \cite{rodighiero04}). \cite{vaisanen02} put this discrepancy down to  calibration issues with the ISOCAM data and corrected the ELAIS-N source counts down by a factor of $\sim$1.75. However, even after this correction, a significant discrepancy remains. In this work, the  {\it AKARI} NEP-wide data appears to be more consistent with the results from the ELAIS-N surveys with the source counts lying significantly above the results from the ELAIS-S fields. The resolution of this discrepancy is an important issue for galaxy evolution modelling which relies on the bright end of the source counts to set the normalization of the models.

\subsection{Comparison with evolutionary models}\label{sec:comparemodels}

\begin{figure*}
\centering
\centerline{
\psfig{ figure=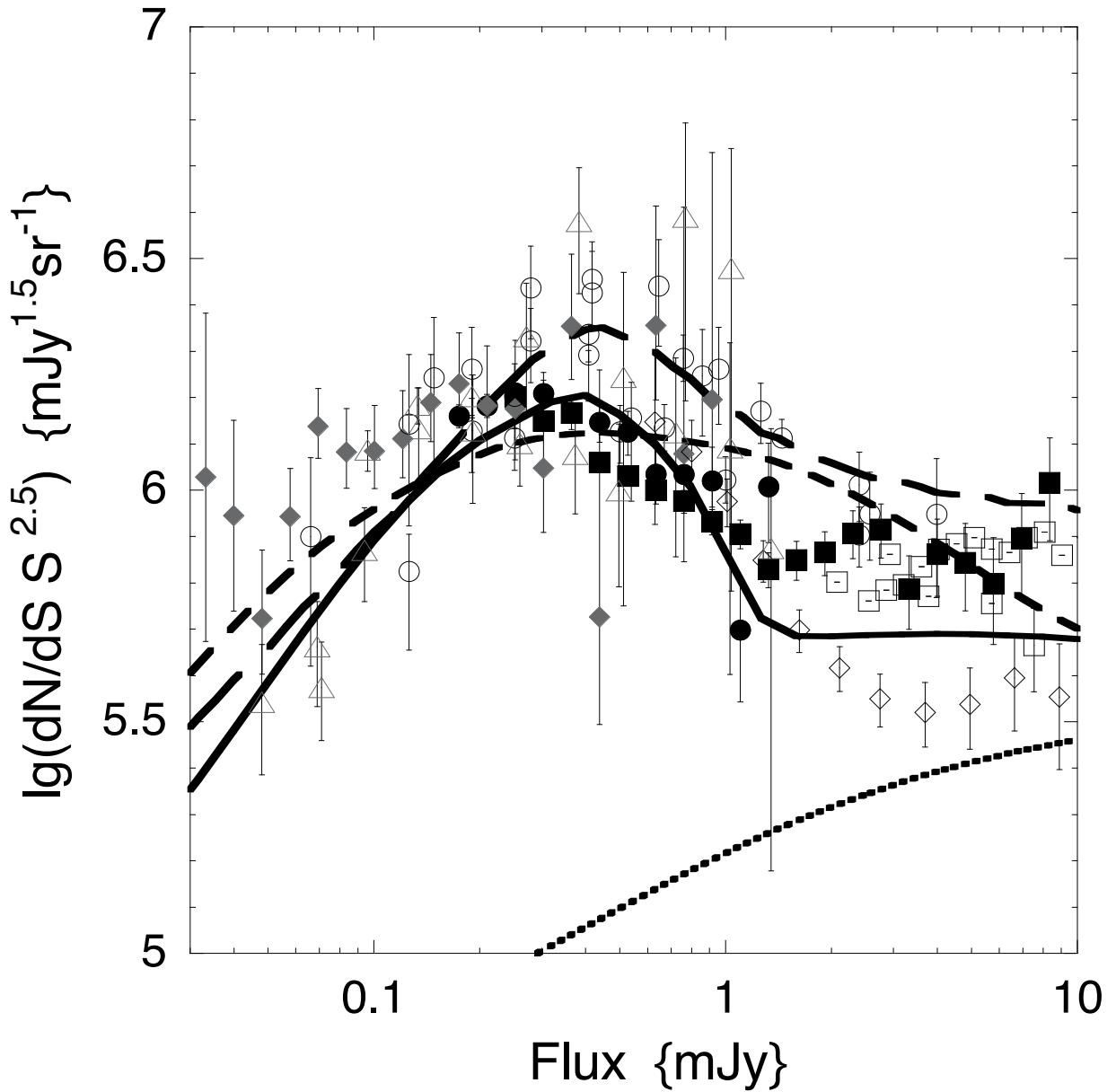,width=16cm}
}
\caption{Comparison of the {\it AKARI}-NEP survey L15 band differential source counts with the observational result of the surveys carried out at 15$\mu$m by {\it ISO} \& {\it Spitzer}. 
The  {\it AKARI} NEP-deep obseervations are shown as {\it filled circles} and the  {\it AKARI} NEP-wide observations as {\it filled squares}. Also plotted are the  15$\mu$m differential source counts from the {\it Spitzer}/IRS GOODS-N \& GOODS-S surveys {\it open triangles}, Teplitz et al. (2005a), Teplitz (2005b); deep {\it ISO} surveys of {\it open circles} Altieri et al. (1999), Oliver et al. (1997), Aussel et al. (1999), Elbaz et al. (1999);   {\it ISO} ELAIS-N  {\it open squares} Serjeant et al. (2000); {\it ISO} ELAIS-S {\it open diamonds} Gruppioni et al (2002).  In addition, for reference we plot the galaxy evolution models of Pearson (2005) as {\it dotted} (no-evolution scenario) and {\it solid} (evolving scenario), the models of Rowan-Robinson (2009)  {\it dashed} line,  and the models of Lagache (2004) as a {\it dot-dash} line .
\label{isospitzercounts}}
\end{figure*}

In  Figure \ref{isospitzercounts} we also compare the observed counts with three contemporary galaxy evolution models. We overplot the evolutionary models of  \cite{lagache04}, \cite{pearson05} \&  \cite{mrr09} along with the non-evolving model of  \cite{pearson05}. 
The evolution in the 15$\mu$m source counts is immediately apparent by comparison with the no-evolution model of \cite{pearson05} which falls off steeply at bright ($S>$10mJy) flux levels.  All of the three evolutionary models fit the location of the peak flux in the 15$\mu$m counts over the 0.2$<S<$0.4mJy range. The \cite{lagache04} model over predicts the {\it AKARI} counts at all fluxes brighter than 0.2mJy and has a high normalization at the bright end. This model does fit the counts from the  {\it ISO} deep surveys, however it should be noted from Table ~\ref{surveys} that the {\it AKARI} NEP-deep Survey covers 1--2 orders of magnitude larger area and should provide a more reliable estimate of the source counts at any given flux level . 
The model of \cite{pearson05} produces a good fits to both the faint ($<$1mJy)  {\it ISO} \& {\it Spitzer} counts and is consistent with the  {\it AKARI} NEP-deep counts down to the 200$\mu$Jy level. At fainter fluxes the predicted source counts fall off faster than the  {\it AKARI}  data. The model has a slightly lower normalization compared to the {\it AKARI} NEP-wide and {\it ISO} ELAIS-N counts at fluxes brighter than 200mJy due to the necessity to strike a compromise between the normalization between the ELAIS-N and ELAIS-S counts. 
Although the \cite{mrr09}  model fits the {\it AKARI} source counts reasonably well at the sub-mJy level, the model misses the upturn in the flux at S$\sim$1-2mJy. 
All three models diverge at the faintest fluxes ($S<$0.1mJy) where the source counts are not well constrained and where the effects of galaxy confusion dominate.

\begin{table*}
\caption{Comparison of  major surveys carried out at 15$\mu$m listed by mission, then by decreasing areal coverage. }
\centering
\begin{tabular}{@{}llllll}
\hline\hline
Satellite &Survey  & Area & Sensitivity & Sources & Volume \\
              &            & (deg.$^{2}$) & (mJy) & (number) & (Mpc$^{3}$)\\
\hline
{\it AKARI } & NEP-wide$^1$ & $5.8$  & $0.25 *$& 10686 & 57.2\\
{\it AKARI } & NEP-deep$^2$ & $0.6$  & $0.1 *$& 6737 & 6.17\\
{\it AKARI } & NEP-PV$^3$ & $0.02$  & $0.07$& 277 & 0.21 \\
{\it AKARI } & Abell 2218$^4$ & $0.034$  & $0.042$& 565 & 0.35 \\
{\it Spitzer } & HDF-S$^5$ & $0.04$ ($0.003$) & $0.09$ ($0.04$) & 515 & 0.41 (0.031)\\
{\it Spitzer } & HDF-N$^6$ & $0.01$ & $0.08$& 153 & 0.1 \\
{\it ISO } & ELAIS S1$^7$ & $4.17 $  & $0.5$& 736 & 38.86\\
{\it ISO } & ELAIS N1$^7$ & $2.67 $ & $0.5$& 490 & 24.88\\
{\it ISO } & ELAIS N2$^7$ & $2.67 $  & $0.5$& 566 & 24.88 \\
{\it ISO } & ELAIS N3$^7$ & $0.88 $  & $0.5$& 131 & 8.2 \\
{\it ISO } & Lockman Hole Shallow$^8$ & $0.55 $ & $0.25$& 457 & 5.41\\
{\it ISO } & Marano Deep$^9$ & $0.25 $  & $0.3$ & 180 & 2.44 \\
{\it ISO } & Lockman Hole Deep$^{10}$ & $0.14$  & $0.2$& 283 & 1.39\\
{\it ISO } & ELAIS S2$^{11}$ & $0.12 $  & $0.4$& 43 & 1.15\\
{\it ISO } & CFRS 1452 Field$^{12}$ & $0.028$ & $0.4$ & 41 & 0.27 \\
{\it ISO } & Marano UDSR$^9$ & $ 0.025$ & $0.3$ & 142 & 0.244\\
{\it ISO } & Marano UDSF$^9$ & $ 0.025$ & $0.3$ & 137 & 0.244\\
{\it ISO }& A370 Cluster$^{13}$ & $0.009$ & $0.2$& 20 & 0.09 \\
{\it ISO } & HDF-S$^{14}$ & $0.008$ & $0.25$ & 24 & 0.079\\
{\it ISO } & HDF-N$^{15}$ & $0.007$ & $0.2$ & 19 & 0.069\\
{\it ISO }& A2218 Cluster$^{13}$ & $0.006$ & $0.1$& 46 & 0.062  \\
{\it ISO }& A2390 Cluster$^{16}$ & $0.0015$  & $0.1$& 34 &  0.015 \\
\hline
\multicolumn{6}{l}{$^*$  5$\sigma$ sensitivity above the noise }\\
\multicolumn{6}{l}{$^1$  \cite{lee08} }\\
\multicolumn{6}{l}{$^2$  \cite{wada08} }\\
\multicolumn{6}{l}{$^3$  \cite{lee08} }\\
\multicolumn{6}{l}{$^4$  \cite{hopwood10} }\\
\multicolumn{6}{l}{$^5$    \cite{teplitz05b}  }\\
\multicolumn{6}{l}{$^6$    \cite{teplitz05a}  }\\
\multicolumn{6}{l}{$^7$  \cite{vaccari05}  }\\
\multicolumn{6}{l}{$^8$  \cite{fadda04}  }\\
\multicolumn{6}{l}{$^9$  \cite{elbaz99}   }\\
\multicolumn{6}{l}{$^{10}$ \cite{rodighiero04}  }\\
\multicolumn{6}{l}{$^{11}$  \cite{pozzi03}  }\\
\multicolumn{6}{l}{$^{12}$   \cite{flores99}   }\\
\multicolumn{6}{l}{$^{13}$  \cite{metcalfe03}  }\\
\multicolumn{6}{l}{$^{14}$  \cite{oliver02}   }\\
\multicolumn{6}{l}{$^{15}$  \cite{serjeant97}  }\\
\multicolumn{6}{l}{$^{16}$  \cite{altieri99} }\\
\end{tabular}
\label{surveys}
\end{table*}

\smallskip

\section{Summary}\label{sec:summary}

We have derived the source counts at 15$\mu$m from {\it AKARI}-IRC L15 band observations of the  {\it AKARI} large-area survey at the North Ecliptic Pole. Source counts from both the NEP-deep survey, (covering a larger area of $\sim$0.6 square degrees compared to the uniform survey area of 0.38 square degrees reported in \cite{wada08}) and the NEP-wide survey, covering 5.8 square degrees are presented. The total number of sources detected are 6737 and 10700 down to 5$\sigma$ limiting fluxes of 117 \& 250$\mu$Jy for the NEP-deep and NEP-wide survey respectively. 

The Euclidean normalized differential source counts  have been shown for both surveys and we have compared the  {\it AKARI} results with the previous surveys carried out at or around 15$\mu$m with the {\it ISO} and {\it Spitzer} observatories. The 15$\mu$m  {\it AKARI} data cover the entire flux range from the deepest to shallowest   {\it ISO} and {\it Spitzer} surveys (40$\mu$Jy-$>$10mJy) over areas sufficiently significant to overcome cosmic variance in the resulting statistics detecting six times as many sources as the widest surveys carried out with  {\it ISO}.

The  {\it AKARI}  data is consistent with the evidence of strong evolution appearing in the source counts at fluxes of around a milli-Jansky and the super-Euclidean slopes observed in the differential sources counts at sub-milliJansky levels. At fainter fluxes the   {\it AKARI}  data is consistent with the  {\it ISO} and {\it Spitzer} source counts down to fluxes of 200$\mu$Jy, reproducing the turnover in the differential source counts around 300$\mu$Jy. However, at the faintest fluxes ($<$200$\mu$Jy), the  {\it AKARI} source counts fall off much slower than observed in the {\it Spitzer}  data or predicted by evolutionary models, although the counts are still marginally consistent with the deep  lensed differential counts from {\it ISO}. At these levels, confusion noise due to crowded fields may be becoming significant, affecting both the source extraction and the completeness corrections, and we estimate a confusion limit of $\sim$130$\mu$Jy. Further processing of the data, including multi-wavelength source counts from the final band merged catalogue will help to determine whether the faint excess is a real effect or an artifact of the source extraction in crowded fields.

At brighter fluxes, the NEP-wide data provides vital information on the bright-end normalization of the source counts at 15$\mu$m, given the uncertainty in the calibration of the  {\it ISO} source counts at fluxes brighter than a milliJansky.
We find that the NEP-wide source counts converge to a flat Euclidean distribution more consistent with the  {\it ISO} ELAIS-N source counts than the ELAIS-S field, even after the calibration corrections suggested by \cite{vaisanen02}.

We have compared the source counts from the {\it AKARI}  NEP survey with three contemporary galaxy evolution models, finding that the \cite{lagache04} model over predicts the {\it AKARI} source counts at all flux levels and the  \cite{mrr09} model misses the upturn of the source counts at 1-2mJy. The models of  \cite{pearson05} fit both the magnitude and position of the  the upturn in the source counts  and the peak between  0.2$<S<$0.4mJy but has a slightly lower normalization at the brightest fluxes compared to the NEP-wide counts. All three models diverge from each other at fainter fluxes where the source counts become confusion dominated.

\smallskip

\begin{acknowledgements}
The authors wish to thank the referee, whose detailed comments greatly improved the focus and quality of this work.\\ HML was supported by National Research Foundation of Kore (NRF) grant No. 2006-341-C00018.\\
The {\it AKARI} Project is an infrared mission of the Japan Space Exploration Agency (JAXA) Institute of Space and Astronautical Science (ISAS), and is carried out with the participation of mainly the following institutes; Nagoya University, The University of Tokyo, National Astronomical Observatory Japan, The European Space Agency (ESA), Imperial College London, University of Sussex, The Open University (UK), University of Groningen / SRON (The Netherlands), Seoul National University (Korea). The far-infrared detectors were developed under collaboration with The National Institute of Information and Communications Technology.
\end{acknowledgements}



\begin{thebibliography}{}
\addcontentsline{toc}{section}{References}

\bibitem[{{Altieri et al. }(1999)}]{altieri99} 
Altieri, B., Metcalfe, L.,  Kneib, J. P. et al., 1999, AA, 343, L65

\bibitem[{{Aussel et al. }(1999)}]{aussel99} 
Aussel, H., Cesarsky, C.J., Elbaz, D. et al., 1999, AA, 342, 313

\bibitem[{{Bertin \&  Arnouts}(1996)}]{bertin96} 
Bertin, E. Arnouts, S. 1996, AAS, 117, 393

\bibitem[{{Blain et al. }(1999a)}]{blain99} 
Blain, A.W.,Kneib, J.-P., Ivison, R.J. et al., 1999a, ApJ, 302, 632

\bibitem[{{Condon }(1974)}]{condon74} 
Condon, J.J., 1974, ApJ, 188, 279

\bibitem[{{Chary et al. }(2004)}]{chary04} 
Chary, R., Casertano, S,  Dickinson, M. E. et al., 2004, ApJSS, 154, 80

\bibitem[{{Dole et al. }(2003)}]{dole03} 
Dole, H., Lagache, G,  Puget J.-L., 2003, ApJ, 585, 617

\bibitem[{{Dole et al. }(2004)}]{dole04} 
Dole, H., Reike, G.H., Lagache, G et al., 2004, ApJ, 154, 93

\bibitem[{{Elbaz et al. }(1999)}]{elbaz99} 
Elbaz, D., Cesarsky, C. J., Fadda, D. et al., 1999, AA, 351, L37

\bibitem[{{Elbaz et al. }(2005)}]{elbaz05} 
Elbaz, D., 2005, Space Science Reviews, 119, 93

\bibitem[{{Fadda et al. }(2004)}]{fadda04} 
Fadda, D. Lari, C., Rodighiero, G. et al.,  2004, AA, 427, 23

\bibitem[{{Flores et al. }(1999)}]{flores99} 
Flores, H., Hammer, F., Thuan, T. X. et al., 1999, ApJ 517, 148

\bibitem[{{Gruppioni et al. }(2002)}]{gruppioni02} 
Gruppioni, C., Lari, C., Pozzi, F. et al., 2002, MNRAS, 335, 831

\bibitem[{{Hogg }(2001)}]{hogg01} 
Hogg, D.W., 2001, AJ, 121, 1207

\bibitem[{{Hopwood }(2010)}]{hopwood10} 
Hopwood, R., Serjeant S., Negrello, M. et al., 2010, ApJ, in press

\bibitem[{{Houck et al. }(2004)}]{houck04} 
Houck, J.R., Roellig, T. L., van Cleve, J. et al., 2004, ApJSS, 154, 18

\bibitem[{{Hughes et al. }(1998)}]{hugh98} 
Hughes, D., Serjeant, S., Dunlop J. et al., 1998, Nature, 394, 241

\bibitem[{{Hwang et al. }(2007)}]{hwang07} 
Hwang, N., Lee, M.G., Lee H.M. et al., 2007, ApJS, 172, 583

\bibitem[{{Jeong et al. }(2006)}]{jeong06} 
Jeong, W.-S., Pearson, C.P., Lee H.M. et al., 2006, MNRAS, 369, 281

\bibitem[{{Kawada et al.  }(2007)}]{kawada07} 
Kawada, M., Baba, H., Barthel, P. D. et al., 2007, PASJ, 59, 389

\bibitem[{{Kessler et al. }(1996)}]{kessler96} 
Kessler, M. F., Steinz, J. A., Anderegg, M. E. et al., 1996, AA, 315, L27

\bibitem[{{Lagache et al. }(2004)}]{lagache04} 
Lagache, G., et al. 2004, ApJS, 154, 112

\bibitem[{{Lee et al. }(2008)}]{lee08} 
Lee, H. M., Kim, S. J., Im, M. et al., 2009, PASJ, 61, 375

\bibitem[{{Lorente et al. }(2007)}]{lorente07} 
Lorente, R., Onaka, T., Ita, Y. et al., 2007, AKARI IRC data user manual, ver.1.3 

\bibitem[{{Magliocchetti et al. }(2007)}]{magliocchetti07} 
Magliocchetti M., Silva L., Lapi A. et al., 2007, MNRAS, 375, 1121

\bibitem[{{Magliocchetti et al. }(2008)}]{magliocchetti08} 
Magliocchetti M., Cirasuolo M., McLure R.J. et al., 2008, MNRAS, 383, 1131

\bibitem[{{Matsuhara et al. }(2006)}]{matsuhara06} 
Matsuhara H., Wada, T., Matsuura, S. et al., 2006, PASJ, 58, 673

\bibitem[{{Metcalfe  et al. }(2003)}]{metcalfe03}
Metcalfe, L., Kneib, J.-P., McBreen, B. et al. , 2003, AA, 407, 791

\bibitem[{{Mortier  et al. }(2005)}]{mortier05}
Mortier A.M.J., Serjeant, S., Dunlop, J. S. et al., 2005, MNRAS, 363, 563

\bibitem[{{Moshir et al. }(1990)}]{moshir90} 
Moshir, M., Kopan, G., Conrow, T. et al., 1990, Infrared Astronomical Satellite Catalog, Faint Source catalog version 2.0

\bibitem[{{Murakami et al. }(2007)}]{murakami07} 
Murakami, H., Baba, H., Barthel, P. et al., 2006, PASJ, 59, 369

\bibitem[{{Oliver et al. }(1997)}]{oliver97} 
Oliver, S. J., Goldschmidt, P., Franceschini, A. et al., 1997, MNRAS, 289, 471

\bibitem[{{Oliver et al. }(2002)}]{oliver02} 
Oliver S.J., Mann R.G., Carballo R. et al., 2002, MNRAS, 289, 471

\bibitem[{{Onaka et al. }(2007)}]{onaka07} 
Onaka, T., Matsuhara, H., Wada, T. et al., 2007, PASJ, 59, 401

\bibitem[{{Papovich et al. }(2004)}]{papovich04} 
Papovich, C., Dole, H., Egami, E. et al.,  2004, ApJSS, 154, 70

\bibitem[{{Pearson }(2005)}]{pearson05} 
Pearson, C.P.,  2005, MNRAS, 358, 1417

\bibitem[{{Pozzi et al. }(2003)}]{pozzi03} 
Pozzi F. et al., 2003, MNRAS, 343, 1348

\bibitem[{{Pozzi et al. }(2004)}]{pozzi04} 
Pozzi, F., Ciliegi, P., Gruppioni, C. et al., 2004, MNRAS, 609, 122

\bibitem[{{Rodighiero et al. }(2004)}]{rodighiero04} 
Rodighiero, G. Lari, C., Fadda, D. et al.,  2004, AA, 427, 773

\bibitem[{{Rodighiero et al. }(2006)}]{rodighiero06} 
Rodighiero, G. Lari, C., Pozzi, F. et al.,  2006, MNRAS, 371, 1891

\bibitem[{{Rowan-Robinson et al. }(2004)}]{mrr04}
Rowan-Robinson, M., Lari, C., Perez-Fournon, I. et al, 2004, MNRAS, 351, 1290

\bibitem[{{Rowan-Robinson et al. }(2009)}]{mrr09}
Rowan-Robinson, M. 2009, MNRAS, 394, 117

\bibitem[{{Rush, Malkan \& Spinoglio }(1993)}]{rush93} 
Rush, B., Malkan, M., Spinoglio, L., 1993, ApJSS, 89, 1

\bibitem[{{Scott et al. }(2002)}]{scott02} 
Scott, S. E. Fox, M. J. Dunlop, J. S.et al., 2002, MNRAS, 331, 817

\bibitem[{{Serjeant et al. }(1997)}]{serjeant97} 
Serjeant, S. B. G., Eaton, N., Oliver, S. J. et al., 1997, MNRAS, 289, 457

\bibitem[{{Serjeant et al. }(2000)}]{serjeant00} 
Serjeant, S., Oliver, S., Rowan-Robinson, M. et al., 2000, MNRAS, 316, 768

\bibitem[{{Smail, Ivison \& Blain }(1997)}]{smail97}
Smail, I., Ivison, R.J., Blain, A.W., 1997, ApJ, 490, L5

\bibitem[{{Soifer et al. }(1987)}]{soifer87} 
Soifer, B.T., Houck, J.R., Neugebauer, G., 1987, ARAA, 25, 187

\bibitem[{{Somerville et al. }(2004)}]{somerville04} 
Somerville, R.S., Lee, K., Ferguson, H.C. et al., 2004, ApJL, 600, L171 

\bibitem[{{Takeuchi et al. }(2001)}]{takeuchi01} 
 Takeuchi, T.T., Ishii, T.T., Hirashita, H. et al., 2008, PASJ, 60, 375

\bibitem[{{Tanabe et al. }(2008)}]{tanabe08} 
 Tanabe, T., Sakon, I., Cohen, M. et al., 2001, PASJ, 53, 37

\bibitem[{{Takagi et al. }(2007)}]{takagi07} 
Takagi, T., Matsuhara, H., Wada, T. et al., 2007, PASJ, 59, 557

\bibitem[{{Teplitz et al. }(2005a)}]{teplitz05a} 
Teplitz, H.I., Charmandaris, V., Chary, R.R. et al., 2005a, ApJ, 634, 128

\bibitem[{{Teplitz et al. }(2005b)}]{teplitz05b} 
Teplitz, H.I. 2005b, in eds D. Elbaz, H. Aussel, ÓWhen UV meets IR: a History of Star FormationÓ, XXVth Moriond Astrophysics Meeting, La Thuile, Italy, March 6-12, 2005

\bibitem[{{Vaccari et al. }(2005)}]{vaccari05} 
Vaccari, M., Lari, C., Angeretti, L. et al., 2005, MNRAS, 358, 397

\bibitem[{{Vaisanen et al. }(2001)}]{vaisanen01} 
Vaisanen, P., Tollestrup, E.V., Fazio, G.G., 2001, MNRAS, 325, 1241

\bibitem[{{Vaisanen et al. }(2002)}]{vaisanen02} 
Vaisanen, P., Morel, T., Rowan-Robinson, M. et al. 2002, MNRAS, 337, 1043

\bibitem[{{Wada et al. }(2007)}]{wada07} 
Wada, T., Oyabu, S., Ita, Y. et al., 2007, PASJ, 59, 515

\bibitem[{{Wada et al. }(2008)}]{wada08} 
Wada, T., Matsuhara, H., Oyabu S. et al., 2007, PASJ, 59, 515

\end{thebibliography}
\end{document}